\documentclass[nonacm,sigplan]{acmart}

\usepackage[]{hyperref}
\usepackage{amsmath}
\usepackage{subfig}
\usepackage{enumitem}
\usepackage{xspace}
\usepackage{multirow}
\usepackage{listings}
\usepackage{xcolor}
\usepackage{soul}
\usepackage{fancyvrb}
\usepackage[export]{adjustbox}
\setlist[itemize]{leftmargin=*}

\definecolor{bg}{rgb}{0.95,0.95,0.95}

\begin{document}
\newcommand{\todo}[1]{\textcolor{blue}{TODO: #1}}
\newcommand{\txs}[1]{{\small\texttt{#1}}}
\newcommand{\txsfig}[1]{{\footnotesize\texttt{#1}}}
\newcommand{\swatch}[1]{\textcolor{#1}{$\blacksquare$}}
\definecolor{deepred}{HTML}{EA2027}
\definecolor{deepblue}{HTML}{1B1464}
\definecolor{forestgreen}{HTML}{006266}
\definecolor{ultraviolet}{HTML}{341f97}
\definecolor{seafoam}{HTML}{1289A7}
\definecolor{radiantyellow}{HTML}{F79F1F}
\definecolor{burntorange}{HTML}{d35400}
\definecolor{darkgrey}{HTML}{273c75}
\definecolor{pastelblue}{HTML}{E3F4F4}

\definecolor{keywordcolor}{HTML}{1289A7}
\definecolor{apicolor}{HTML}{EA2027}
\definecolor{functioncolor}{HTML}{1B1464}
\definecolor{variablecolor}{HTML}{341f97}
\definecolor{argumentcolor}{HTML}{006266}
\definecolor{commentcolor}{HTML}{F79F1F}

\definecolor{matplotlibgreen}{HTML}{008000}

\newcommand{\tool}{\textsf{Shem}\xspace{}}
\newcommand{\specabbrev}{XXXSpec}
\newcommand{\ark}{Ark}
\newcommand{\normaldist}{\mathcal{N}}
\newcommand{\codein}[1]{{\texttt{\small #1}}}
\newcommand{\mathin}[1]{{\small\textbf{$#1$}}}
\newcommand{\rulein}[1]{{\small\textsc{#1}}}

\newcommand{\jax}{JAX}
\newcommand{\equinox}{Equinox}

\newcommand{\ddt}[1]{\frac{d#1}{dt}}
\newcommand{\noiseamp}{\alpha}
\newcommand{\loss}{\mathcal{L}}
\newcommand{\vecstatevar}{\mathbf{x}}
\newcommand{\param}{\theta}
\newcommand{\vecparam}{\boldsymbol{\param}}
\newcommand{\vecadjoint}{\mathbf{a}}
\newcommand{\vecadjointgrad}{\mathbf{u}}

\newcommand{\vecmismatch}{\boldsymbol{\delta}}
\newcommand{\vectransient}{\boldsymbol{\xi}}

\newcommand{\cnn}{CNN}
\newcommand{\obc}{OBC}
\newcommand{\tln}{TLN}
\newcommand{\nn}{NNs}

\newcommand{\nacp}{NACP}

\newcommand{\tightbullet}{\noindent}
\newcommand{\proseheading}[1]{\tightbullet{\textbf{{#1}}}}
\newcommand{\bulletkeyword}[1]{\textbf{#1}}
\newcommand{\sectionbullet}[2]{\textbf{#1[{\small\texttt{#2}}].}}
\newcommand{\mathbullet}[2]{\textbf{#1[{\small$#2$}].}}

\newcommand{\itoo}{\overline{I2O}}

\newcommand{\allcplweight}{k}
\newcommand{\cplweight}{\allcplweight_{ij}}
\newcommand{\lockweight}{l}

\newcommand{\makeset}[1]{\mathbf{#1}}
\newcommand{\makefunc}[1]{\mathrm{#1}}
\newcommand{\bit}[1]{b_{#1}}
\newcommand{\chl}{c}
\newcommand{\centerchls}{\makeset{C}}

\newcommand{\PUFrawfn}{\makefunc{D}}
\newcommand{\PUFbinfn}{\makefunc{P}}
\newcommand{\bfneighbor}{\makefunc{bf}}
\newcommand{\sign}{\makefunc{is\_positive}}

\title{\tool{}: A Hardware-Aware Optimization Framework for Analog Computing Systems}

\author{Yu-Neng Wang}
\email{wynwyn@stanford.edu}
\affiliation{%
  \institution{Stanford University}
  \country{USA}
}
\author{Sara Achour}
\email{sachour@stanford.edu}
\affiliation{%
  \institution{Stanford University}
  \country{USA}
}

\begin{abstract}
As the demand for efficient data processing escalates, reconfigurable analog hardware which implements novel analog compute paradigms, is promising for energy-efficient computing at the sensing and actuation boundaries. These analog computing platforms embed information in physical properties and then use the physics of materials, devices, and circuits to perform computation. These hardware platforms are more sensitive to nonidealities, such as noise and fabrication variations, than their digital counterparts and accrue high resource costs when programmable elements are introduced. Identifying resource-efficient analog system designs that mitigate these nonidealities is done manually today. 

While design optimization frameworks have been enormously successful in other fields, such as photonics, they typically either target linear dynamical systems that have closed-form solutions or target a specific differential equation system and then derive the solution through hand analysis. In both cases, time-domain simulation is no longer needed to predict hardware behavior. In contrast, described analog hardware platforms have nonlinear time-evolving dynamics that vary substantially from design to design, lack closed-form solutions, and require the optimizer to consider time explicitly.  We present \tool{}, an optimization framework for analog systems. \tool{} leverages differentiation methods recently popularized to train neural ODEs to enable the optimization of analog systems that exhibit nonlinear dynamics, noise and mismatch, and discrete behavior. We evaluate \tool{} on oscillator-based pattern recognizer, CNN edge detector, and transmission-line security primitive design case studies and demonstrate it can improve designs. To our knowledge, the latter two design problems have not been optimized with automated methods before.

\end{abstract}

\maketitle 
\pagestyle{plain} 

\section{Introduction}

There has been an emergence of new workloads that place extreme power constraints on the hardware and require processing near the sensing and actuation interfaces~\cite{lequepeys2021overcoming,outeiral2021prospects, bayerstadler2021industry,irimia2012green}. Analog computing systems are a promising class of hardware that can perform processing directly on analog signals domain and often at very low energy, enabling processing large amounts of analog data with little digitization~\cite{Decadal, murmann2020a2i}. 
Analog computing systems encode information in physical signals (e.g., voltage) and then leverage the dynamics of materials, devices, and circuits to perform computation. 
Modern analog computing systems are reconfigurable and implement novel analog computational paradigms, such as oscillator-based computing and cellular nonlinear networks, which are inherently nonlinear and often use non-standard physical properties~\cite{ryynanen2001dual,gangopadhyay2014compressed,mehonic2020memristors,konatham2020real,sebastian2020memory}. Ordinary differential equations capture the semantics of the analog compute paradigm and executing computations involves solving or simulating the differential equation system.

\proseheading{Design Challenges.} Identifying a resource-efficient, system-level design for this class of analog hardware that delivers acceptable fidelity remains a significant challenge and is primarily performed manually. First, analog hardware is sensitive to \textit{nonidealities} that affect the fidelity of the computation, such as noise, fabrication-induced parameter variations, and environmental sensitivities. Second, the digital interface circuitry used to program the hardware and perform digital/analog conversion increases resource usage, especially with increasing precision~\cite{cowan2005vlsi, huang2017hybrid, tsividis2018analog-computer, guo2016hybrid-computer,achour2020Legno}. Reducing the precision of these digital elements and the degree of programmability significantly reduces the complexity of the design. \textit{Enabling automated optimization of these analog systems would enable identification of useful design.}

\subsection{Existing Design Optimization Methods} 

Design optimization tools find design parameterizations that minimize some \textit{cost function}, which captures the end-to-end system-level property the designer desires. Gradient-based optimization methods have been extensively used in photonics~\cite{hughes2018adjoint, su2020nanophotonic, molesky2018inverse, li2022physics, li2023lightridge, molesky2018inverse}. Because these optimizers use gradient information to drive the search intelligently and are built on heavily accelerated ML frameworks, designs of tens or even hundreds of thousands of design variables can be targeted~\cite{kang2024large, piggott2020inverse}. Gradient-based optimizers have also been used to optimize analog circuits to a lesser degree; these methods optimize SPICE circuits to minimize specific circuit metrics (e.g., delay) and typically leverage optimization techniques that specifically work with circuit schematics~\cite{ rohrer1967fully,director1969generalized, conn1998jiffytune, conn1999gradient, visweswariah2000noise, joshi2017analog, hu2020adjoint, li2023circuit}. 

\proseheading{Limitations.} The gradient-based methods devised in these prior works are insufficient for optimizing analog computing systems. To be able to use these methods, the gradient of the cost function must be taken. Previously developed optimizers primarily target linear dynamical systems that have closed-form solutions or target a specific differential equation system that has a hand-derived solution. In both cases, time-domain simulation is no longer needed to predict hardware behavior, and the gradient is relatively straightforward to compute. In contrast, this class of analog systems has nonlinear dynamics that rarely have analytic solutions, so gradient needs to be taken over the time domain simulation of the system. 

Analog systems also experience nonidealities, such as fabrication-induced errors and noise that introduce stochasticity into the system's dynamics. This stochastic behavior is not inherently differentiable and interacts with the system's nonlinearities, producing complex behaviors. In addition, digital logic that exists at the programming and measurement interfaces of the analog system is inherently discrete and, therefore, also not readily differentiable. These behaviors make the system even more challenging to differentiate.

\subsection{Optimization of Analog Systems with \tool{}}

We present \tool{}, an optimization framework for analog systems that directly optimizes over time-domain differential equation models with nonlinear dynamics. \tool{} leverages the adjoint method, a differentiation method recently popularized to train neural ODEs, to directly differentiate over ordinary differential equation simulations and find the gradient of the cost function~\cite{kidger2022neural, chen2018neural, li2020scalable}. \tool{} deploys a translation pass that makes noise, mismatch, and digital logic differentiable so the gradient may be taken. 

We build \tool{} on the \jax{} machine learning framework, which supports auto-differentiation and offers highly optimized backends for auto-parallelization and hardware acceleration, enabling scalable design optimization~\cite{jax2018github}. \tool{} inherits the programming conveniences offered by \jax{}, enabling \tool{} to target a range of analog system designs and optimize complex cost functions that evaluate system-level properties such as end-to-end error.

\subsubsection{Contributions}

\begin{itemize}[leftmargin=*, itemsep=0.1em, parsep=0.1em, topsep=0.2em]
\item{}We introduce techniques that enable time-domain design optimization of analog systems and differentiation over noise, fabrication variations, and digital logic.
\item{}We present \tool{}, a framework built on \jax{} that supports auto-differentiation and optimization of nonlinear time-domain analog systems.
\item{}We evaluate \tool{} on oscillator-based pattern recognizer, cellular nonlinear network edge detector, and transmission-line security primitive design case studies and demonstrate it can improve designs. To our knowledge, the latter two design problems have not been optimized with automated methods before.

\end{itemize}

\section{Example: CNN Edge Detector}\label{sec:example}

\begin{figure}
    \subfloat[CNN Architecture]{
        \centering
        \includegraphics[width=0.7\linewidth]{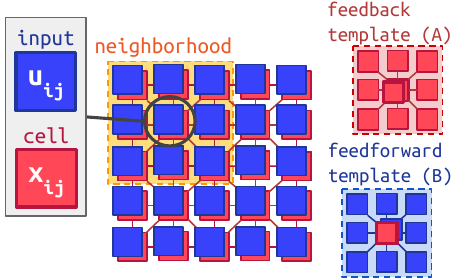}
        \label{fig:cnn:arch}
    }

    \subfloat[Input]{
        \centering
        \includegraphics[height=0.1\textwidth]{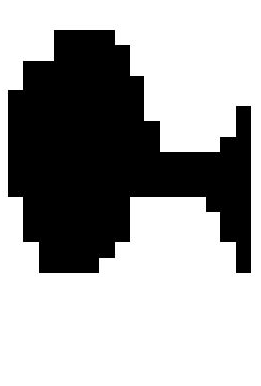}
        \label{fig:cnn:evol:initial}
    }\quad
    \subfloat[Time evolution of CNN]{
        \centering
        \includegraphics[height=0.1\textwidth]{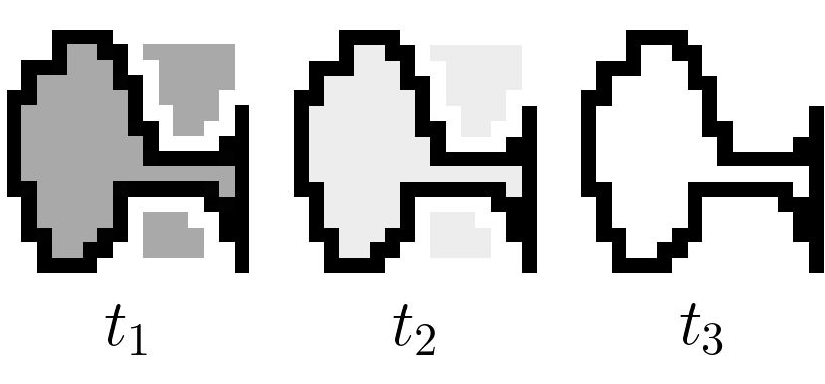}
        \label{fig:cnn:evol:ideal}
    }\quad
    \subfloat[Output]{
        \centering
        \includegraphics[height=0.1\textwidth]{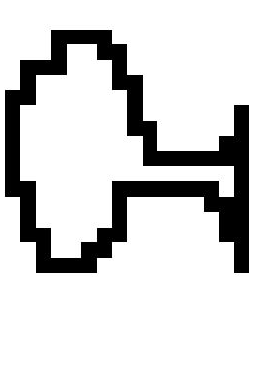}
        \label{fig:cnn:evol:output}
    }

    \subfloat[Dynamical System, $m=1+i-k, n=1+j-l$]{
    \small
    $\ddt{x_{ij}} = -x_{ij} + \sum\limits_{k,l\in \text{Neigh}(i, j)} (A_{m,n}\cdot f(x_{kl}) + B_{m,n}\cdot u_{kl}) + z$
    \label{eq:cnn:dynamics}
    }

    \subfloat[Parameters for CNN Edge Detector]{
        \small
        $A = \begin{bmatrix} 0 & 0 & 0 \\ 0 & 2 & 0 \\ 0 & 0 & 0 \end{bmatrix}, \quad B = \begin{bmatrix} -1 & -1 & -1 \\ -1 & 8 & -1 \\ -1 & -1 & -1 \end{bmatrix}, \quad z = -0.5$
        \label{eq:cnn:template}
    }\hfill
    
    \subfloat[\cnn{} Cell Output Dynamics]{
        \centering
        \includegraphics[width=0.8\linewidth]{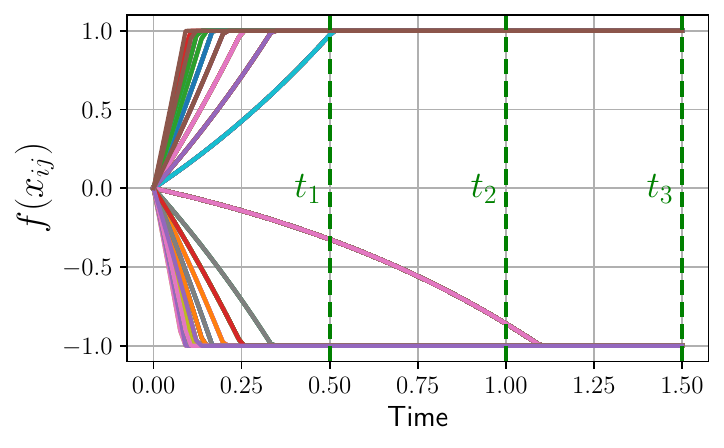}
        \label{fig:cnn:evol:one-pixel}
    }

    \caption{Cellular Nonlinear Network Edge Detector.}
\end{figure}

\begin{figure}
    \centering
    \subfloat[Dynamical System, $m=1+i-k, n=1+j-l$]{
    \small
    $
    \begin{aligned}
        \ddt{x_{ij}} = -x_{ij} + & \sum\limits_{k,l\in \text{Neigh}(i, j)} (\delta_{i,j,k,l}^A\cdot A_{m,n}\cdot f(x_{kl}) + \\ 
        & \delta_{i,j,k,l}^B\cdot B_{m,n}\cdot u_{kl}) + \delta_{i,j}^z\cdot z
    \end{aligned}
    $
    \label{eq:cnn:dynamics-mismatched}
    }

    \subfloat[No Optimization ($\mathit{loss}=0.555$).]{
        \centering
        \includegraphics[width=0.45\linewidth]{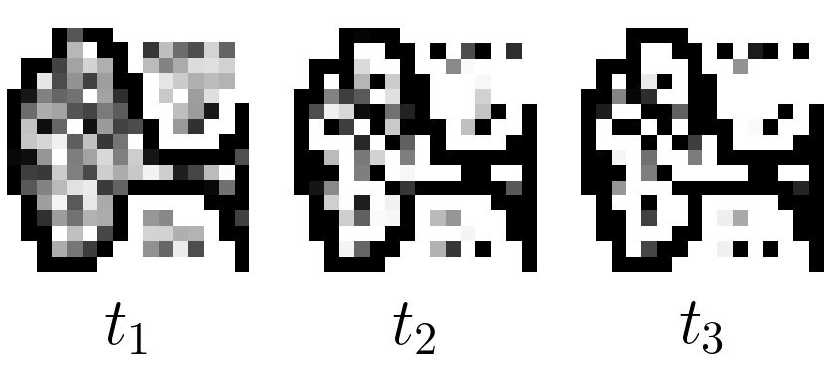}
        \label{fig:cnn:evol:no-opt}
    }\hfill%
    \subfloat[W/ Optimization ($\mathit{loss}=0.032$).]{
        \centering
        \includegraphics[width=0.45\linewidth]{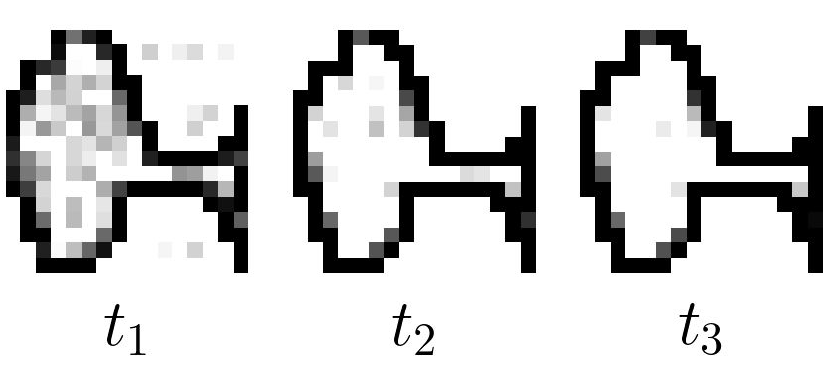}
        \label{fig:cnn:evol:opt}
    }

    \subfloat[t][Loss over Optimizer Iterations]{
        \centering
        \includegraphics[width=0.5\linewidth, valign=b]{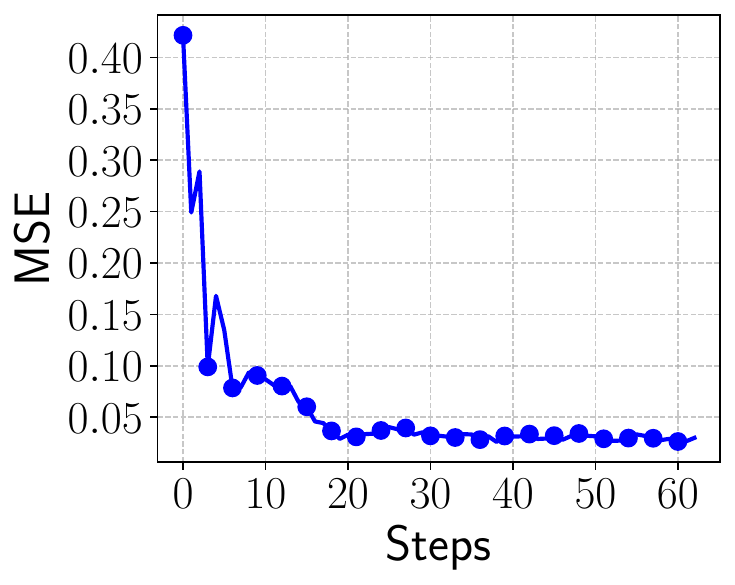}
        \label{fig:eval:cnn:loss-vs-iter}
    }%
    \subfloat[t][Optimized Parameters]{
    \adjustbox{valign=b}{
        \small
        $\begin{aligned}
        & A = \begin{bmatrix} -0.37 & 0.68 & -0.37 \\ 0.68 & 0.60 & 0.68 \\ -0.37 & 0.68 & -0.37 \end{bmatrix}\\
        & B = \begin{bmatrix} -1.21 & -0.68 & -1.21 \\ -0.68 & 7.76 & -0.68 \\ -1.21 & -0.68 & -1.21 \end{bmatrix}\\
        & z = -1.77.\end{aligned}$
        \label{eq:cnn:template-optimized}
    }
    }
    
    \caption{Optimizing CNN edge detector with device mismatch. Initial loss 0.042, optimized loss 0.027.}
    \label{fig:cnn:evol}
\end{figure}


Let us first consider a cellular nonlinear network-based implementation of an edge detector. A cellular nonlinear network (\cnn{}) is an analog compute paradigm that has applications in image processing, partial differential equation (PDE) solving, and security primitives~\cite{chua1988cellular-theory, chua1995cellular-pde, fortuna2001cellular-processing, csaba2009cnn-puf}. 


\proseheading{Overview.} A \cnn{} consists of a grid of locally interconnected cells with values that evolve over time. Figure~\ref{fig:cnn:arch} presents a visualization of the CNN; the nodes map to cells, and edges capture interactions between cells. Each cell at row $i$ column $j$ has a \textit{state variable}  $x_{ij}$ that evolves over time, associated external input $u_{ij}$, and is connected to a neighborhood of cells. In the CNN edge detector application, each external input $u_{ij}$ maps to the greyscale pixel value at position $i,j$ of the image input, and the state variable $x_{i,j}$ settles to the edge detector value. Figure~\ref{fig:cnn:evol:initial} shows the input image provided with external inputs $u_{i,j}$, and Figures~\ref{fig:cnn:evol:ideal} and ~\ref{fig:cnn:evol:output}, respectively, show the evolution of cell outputs over time and the final cell output value. We can see from Figure~\ref{fig:cnn:evol:output} that the CNN edge detector cells on the borders of the shape settle to black.

\proseheading{Differential Equation Model.}  The dynamics of each state variable $x_{ij}$ is modeled with an ordinary differential equation (ODE), which captures the evolution of the cell's value over time. Figure~\ref{eq:cnn:dynamics} presents the ordinary differential equation for cell $x_{ij}$ of the CNN. The derivative of each cell's state $x_{ij}$ is determined by the states of other cells $x$ and external inputs $u$. The state of each cell undergoes a nonlinear activation function $f$. In \cnn{} edge detection, the activation function $f$ clamps the input to the range $\left[-1,1\right]$.

\proseheading{Parameters.} The parameter configuration programs the \cnn{} to perform the edge detection.  The feedback template $A$ determines how the neighboring cells' values contribute to a given cell, and the feedforward template $B$ determines how the neighboring external input values contribute to a given cell. The parameter $z$ offsets the cell's value. Figure~\ref{eq:cnn:template} presents the $A$ template, $B$ template, and $z$ parameter instantiation for the CNN edge detector~\cite{chua1988cellular-app, li2011edge, duan2014memristor}. 

\proseheading{Simulation.} We obtain the output of the CNN edge detector by simulating the differential equations that govern the system over time. Figure~\ref{fig:cnn:evol:one-pixel} shows the cell's output traces over time. Each external input $u_{i,j}$ is set to the pixel values at $i,j$, the cells $x_{i,j}$ are instantiated to zero, and then the dynamical system is simulated over time using a differential equation solver. The edge detector pixel value at a pixel location $i,j$ is measured from the \cnn{} by applying the activation function $f(.)$ to each cell's value. The \cnn{} dynamics have a transient phase of execution where the output pixel values are in flux ($t_1$, $t_2$) and a steady-state phase of execution where output pixels have settled ($t_3$). The result is obtained by measuring the output pixels at steady-state.




\subsection{\cnn{} Edge Detector with Mismatched Analog Circuit}

Cellular nonlinear networks are efficiently implementable with analog hardware as demonstrated in previous studies~\cite{cruz199816cnn-chip, duan2014memristor}. These analog accelerators directly implement the CNN differential equations by leveraging the physics of the underlying circuits. Because circuits are subject to a range of nonidealities, the behavior of the \cnn{} in hardware may deviate substantially from its expected behavior. 

Device mismatch is one such nonideality that has been shown to degrade \cnn{} convergence and edge detection quality in prior works~\cite{fernandez2009cnn-mismatch, wang2024design}. Mismatch occurs when analog hardware experiences random variations during fabrication, which prevent precise control of parameter values and cause identically configured values to deviate slightly between devices~\cite{kinget2005device}. In \cnn{}'s context, mismatch causes deviations in the programmed parameters $A$, $B$, and $z$, making every cell to undergo a slightly different dynamics as shown in Figure~\ref{eq:cnn:dynamics-mismatched} where the $\delta$ random variables modeling the mismatch.


Figure~\ref{fig:cnn:evol:no-opt} shows the time evolution of the CNN edge detector when device mismatch is applied to the $A$, $B$, and $z$ parameters. In the mismatched CNN edge detector, the parameters are drawn from normal distributions centered around the nominal parameter values, with a relative standard deviation of 10\% of the nominal value. That is, $\delta$ are drawn from $\mathcal{N}(1, 0.1)$. This statistical model is commonly used in design to assess the effect of mismatch on the system. Observe that the final output ($t_3$) is much noisier and significantly diverges from the ideal edge detection outcome presented in Figure~\ref{fig:cnn:evol:ideal}. This deviation in behavior occurs because the edge detector template design does not take mismatch into account~\cite{chua1988cellular-theory, chua1988cellular-app, li2011edge, duan2014memristor}. \textit{Can we learn templates for the CNN edge detector that are resilient to device mismatch?}

\subsection{Learning a Mismatch-Aware CNN Edge Detector}

We want to learn a set of CNN parameters that minimize the mean-square error (MSE) between the reference edge detector output $Y$ and the mismatched CNN output at time $t_3$. The following cost function calculates the MSE of pixels between the reference output and output measured from the CNN at time $t_3$:

{\small
\begin{equation}
    \text{MSE}(Y,A,B,z) = \frac{1}{N}\sum_{ij} (y_{ij} - f(x_{ij}[A,B,z](t_{3}))^2.
\end{equation}
}

Because we want to learn a template resilient to mismatch that robustly performs the edge detection task, we want to minimize the \textit{expected} mean-squared error over samples of random mismatch $\delta$ (Figure~\ref{eq:cnn:dynamics-mismatched}). Therefore, the loss function for the mismatch-aware \cnn{} edge detector is

{\small
\begin{equation}
    \text{Loss}(Y,A,B,z) = \mathbb{E}_{\vecmismatch}\left[\text{MSE}(Y,A,B,z)\right].
\end{equation}}

Designers would like to find a set of $A$, $B$, and $z$ that minimizes this expected loss.


\proseheading{Challenges.} 
Minimizing this loss function is highly non-trivial.
Designers may attempt to analytically calculate a closed-form solution for the \cnn{}'s output $f(x_{ij}(t_3))$ to avoid using a differential equation solver entirely. In the case of linear systems, this is possible, but because this system of differential equations is nonlinear, a closed-form solution cannot be easily derived analytically from the differential equations. Even for the \cnn{}, a relatively simple nonlinear dynamical system, analysis is challenging~\cite{chua1988cellular-theory}. These  challenges are further complicated by random mismatches which introduce unpredictable deviations from the ideal behavior.

\begin{figure}[t]
    \centering
    \subfloat[Trainable \cnn{} Model Instantiation]{
    \begin{minipage}{\linewidth}
        \input{figures/cnn-trainable-A-template}
    \end{minipage}
    \label{fig:cnn:A-trainable-snippet}
    }\\
    \subfloat[\cnn{} Edge Detection Optimization Setup. In practice, the for-loop in Lines 16-18 is executed in a vectorized form.]{
    \begin{minipage}{\linewidth}
        \input{figures/cnn-opt-setup}
    \end{minipage}
    \label{fig:cnn:opt-snippet}
    }
\caption{Simplified \tool{} Code Snippets.}
\label{fig:CNN:snippet}
\end{figure}

\subsection{Mismatch-Aware \cnn{} Edge Detection with \tool{}.}

We present \tool{}, a framework that enables the gradient-based optimization of analog computations in the presence of hardware nonidealities. \tool{} supports direct optimization over the time-domain differential equation models of the analog system, enabling optimization over nonlinear dynamical systems that may exhibit stochastic behavior. 

\tool{} works with a \textit{tunable} differential equation model of the dynamical system and a cost function defined over the dynamical system's state variable trajectories. Figure~\ref{fig:cnn:A-trainable-snippet} presents the experimental setup for optimizing the mismatched CNN, written in \tool{}'s API. 

\proseheading{Parameters.} \tool{} allows annotation of the ``trainable'' parameters in the computation. The trainable parameters are those that designers want the optimizer to tune. We leverage the rotation invariance of an edge, leading to symmetric $A$ and $B$ templates to constrain the search space. Lines 2-4 instantiate trainable corner, edge, and center parameters, and Lines 5-7 construct the $A$ template from these parameters. Lines 8-14 do the same for the $B$ template and $z$. With these constraints, the optimizer tunes a total of 7 parameters. The \tool{} API accepts the \cnn{} differential equations and compiles to a differentiable model in Line 19.

\proseheading{Cost Function.} Figure~\ref{fig:cnn:opt-snippet} outlines the optimization setup. The cost function is defined in Lines 1-7. The model set the input $u$ with an image and solves the differential equations in the time domain with sampled random mismatches. The MSE loss is computed over the model readout with a reference edge image. The main optimization loop repeats the cost function to calculate the expected MSE.

\proseheading{Optimization Loop.} Line 10 prepares the training set from the Caltech Silhouettes dataset, which provides $16 \times 16$ black-and-white images~\cite{pmlr-v9-marlin10a}. We set the batch size to 128 . A differentiable \cnn{} model is instantiated in Lines 11-12 with a specified readout time $t_3$. The Adam optimizer, initialized with a learning rate of 0.1, is invoked in Line 13 to enable gradient-based updates~\cite{kingma2017adam}. 
The loop in Lines 14-21 processes the dataset (over 64 steps). MSE losses are computed over each image and its corresponding reference edge (Lines 17-18), and these individual losses are averaged at the end of each batch to obtain the expected MSE loss (Line 19). Based on this loss, Lines 20-21 compute the gradient, allowing the Adam optimizer to adjust the model's trainable weights via gradient descent.

\proseheading{Results.} The optimizer takes a total of 33 minutes on an Intel(R) Xeon(R) Silver 4216 CPU with a Quadro RTX 6000 GPU, totaling 31 seconds per iteration, and optimizes a nonlinear differential equation system containing 256 differential equations over 128 simulations at each iteration. 
The MSE losses over optimization iterations are shown in Figure~\ref{fig:eval:cnn:loss-vs-iter}.

Figure~\ref{fig:cnn:evol:opt} shows the operation of the mismatched-optimized edge detector on the example image after 64 iterations of the gradient descent algorithm, and Figure~\ref{eq:cnn:template-optimized} presents the learned parameters. As we can see after optimization, the final output image looks much cleaner, even though there is a large amount of mismatch in the system. We verified the learned templates on the testing set of the Caltech Silhouettes dataset with distinct random seeds. The optimizer achieves an MSE loss of 0.027 compared to 0.130 from the initial parameters. The results show how \tool{}'s differentiable formulation of the analog system enables the optimization of the \cnn{} edge detector to be robust to device mismatch.

\section{Time-Domain Optimization of Analog Systems}
\label{sec:opt-time-domain}

It is crucial that both analog computation and hardware nonidealities can be represented in differentiable forms. This allows for nonideality-aware optimization using gradient-based methods. In this section, we describe how analog computation can be modeled in a differentiable manner, with techniques from machine learning and scientific computing~\cite{kidger2022neural,li2020scalable,chen2018neural, innes2019differentiable, hindmarsh2005sundials}. 

\subsection{Optimization Problem}

Analog computation can be modeled as a system of ordinary differential equations (ODEs) in the form

{\small
\begin{equation}
    \frac{d\vecstatevar(t)}{dt} = f(\vecstatevar(t), \vecparam, t),
\end{equation}
}

where $\vecstatevar(t) \in \mathbb{R}^n$ represents the state variables at time $t$ with initial condition $\vecstatevar(0)=\vecstatevar_0$. $\vecparam \in \mathbb{R}^m$ denotes the model parameters. $f: \mathbb{R}^n\times\mathbb{R}^m\times\mathbb{R} \rightarrow \mathbb{R}^n$ is the time derivative function of $\vecstatevar$ and is assumed to be differentiable with respect to both $\vecstatevar$ and $\vecparam$. We consider a loss function $\mathcal{L} = \mathcal{L}(\vecstatevar(t_0),\ldots,\vecstatevar(t_k))$ evaluated with the state variables observed at time $t_0,\ldots,t_k$.

\proseheading{Gradient Computation.} We seek to compute the gradient $\frac{d \mathcal{L}}{d \vecparam}$ with respect to the system parameters $\vecparam$, so that we can apply gradient descent-based approaches to optimize the loss function $\mathcal{L}$. This quantity is also referred to as sensitivity in literature~\cite{hindmarsh2005sundials}. Computing the gradient involves tracking the parameters' effect throughout the dynamical system's time evolution and can be obtained with two methods:

\begin{itemize}
\item{}\textit{Backpropagation through Solver} Autodifferentiate a differential equation solver to obtain the gradient with respect to the parameters. More memory intensive.
\item{}\textit{Solving the Continuous Adjoint Equations} Augment the dynamical system with \textit{adjoint} differential equations whose solution is the gradient. More compute intensive.
\end{itemize}

\subsection{Backpropagation through the Solver}
The first approach is to compute the gradient with backpropagation, similar to in a neural network, leveraging the differentiability of numerical ODE solvers.
Numerical solvers compute the ODE solution in discrete time steps $\vecstatevar_{\tau_0}, \vecstatevar_{\tau_1}, \ldots, \vecstatevar_{\tau_N}$, where $\tau_N = t_k$. The state variable at time $\tau_{i}$ is calculated from the previous state(s) and the derivative function, for example, with Euler's method.

{\small
\begin{equation}
    \vecstatevar_{\tau_{i+1}} = \vecstatevar_{\tau_{i}} + \Delta t \cdot f(\vecstatevar_{\tau_{i}}, \vecparam, \tau_{i}),
\end{equation}}
where $\Delta t$ is the time step size. Since $f$ and the update operators are differentiable, $\vecstatevar_{\tau_{i+1}}$ remain differentiable with respect to previous state $\vecstatevar_{\tau_i}$ and $\vecparam$. Therefore, we can apply backpropagation through the solver to compute $\frac{d \mathcal{L}}{d \vecparam}$ with
{\small
\begin{align}
    \begin{split}
        \frac{d \mathcal{L}}{d \vecparam} & = \frac{\partial \mathcal{L}}{\partial \vecstatevar_{\tau_N}}^\top\left(\frac{\partial \vecstatevar_{\tau_N}}{\partial \vecstatevar_{\tau_{N-1}}}\frac{d \vecstatevar_{\tau_{N-1}}}{d \vecparam} + \frac{\partial \vecstatevar_{\tau_{N}}}{\partial \vecparam}\right) \\
        & = \frac{\partial \mathcal{L}}{\partial \vecstatevar_{\tau_N}}^\top\left(\frac{\partial \vecstatevar_{\tau_N}}{\partial \vecstatevar_{\tau_{N-1}}}\left(\frac{\partial \vecstatevar_{\tau_{N-1}}}{\partial \vecstatevar_{\tau_{N-2}}}\frac{d \vecstatevar_{\tau_{N-2}}}{d \vecparam} + \frac{\partial \vecstatevar_{\tau_{N-1}}}{\partial \vecparam}\right) + \frac{\partial \vecstatevar_{\tau_{N}}}{\partial \vecparam}\right) \\
        & \cdots.
    \end{split}
\end{align}
}
Higher-order methods such as the Runge-Kutta method use more previous states to calculate $x_{tau_{i+1}}$ for better accuracy, and the operations still preserve differentiability.

This method is straightforward and efficient with modern automatic differentiation (AD) tools~\cite{jax2018github, kidger2022neural, innes2019differentiable}. The downside is that it can be computationally expensive in terms of memory, as intermediate states and their corresponding gradients need to be stored to compute the loss gradients and scales with the number of time steps.

\subsection{Solving the Continuous Adjoint Equations}

The continuous adjoint method formulates the gradient computation as solving an augmented ODE system, which reduces memory usage by avoiding the need to store all intermediate states~\cite{kidger2022neural,chen2018neural,ma2021adjointcomparison}. The adjoint is defined as $\vecadjoint(t) \equiv \frac{\partial \mathcal{L}}{\partial \vecstatevar(t)}$ and calculates the gradient with the following system of augmented ODEs:

{\small
\begin{equation}
\left\{
\begin{aligned}
    \frac{d\vecstatevar(t)}{dt} &= f(\vecstatevar(t), \vecparam, t), \quad &\vecstatevar(t_k) &= \vecstatevar_{t_k}, \\
    \frac{d\vecadjoint(t)}{dt} &= -\vecadjoint(t)^\top \frac{\partial f(\vecstatevar(t), \vecparam, t)}{\partial \vecstatevar(t)}, \quad &\vecadjoint(t_k) &= \frac{\partial \mathcal{L}}{\partial \vecstatevar(t_k)}, \\
    \frac{d\vecadjointgrad(t)}{dt} &= -\vecadjoint(t)^\top \frac{\partial f(\vecstatevar(t), \vecparam, t)}{\partial \vecparam}, \quad &\vecadjointgrad(t_k) &= 0.
\end{aligned}
\right.
\end{equation}
}

Here, $\vecadjointgrad(0)$ will represent the gradient of interest, $\frac{d \mathcal{L}}{d \vecparam}$. (For a detailed derivation, readers can refer to~\cite{kidger2022neural,chen2018neural}). The initial conditions are set to the state value at the end time $t_k$, which is computed during the forward pass. To obtain $\vecadjointgrad(0)$, we solve the augmented ODEs backward in time, i.e., from $t_k$ to $0$. All state variables in the augmented systems are computed on the fly, thus making the memory complexity independent of the number of time steps.

The key advantage of using the continuous adjoint method is its memory efficiency. However, this approach has drawbacks, including increased runtime and susceptibility to numerical error due to the additional ODE solving. Furthermore, the augmented system may exhibit instability. Therefore, the choice of technique should depend on the characteristics of the specific ODE being solved~\cite{kidger2022neural, ma2021adjointcomparison}.

\section{Differentiable Modeling of Nonidealities}
\label{sec:diff-model-nonideal}

In the basic problem setup, $f(\vecstatevar(t), \vecparam, t)$ is assumed to be differentiable. In practice, analog nonidealities introduce stochastic and non-differentiable behaviors into the dynamics. In addition, digitally programmable elements (e.g., digital/analog converters) and measurement circuits discretize continuous signals, introducing discontinuities. We now extend the basic time-domain gradient calculation method to support common hardware nonidealities.

\subsection{Device Mismatch} 
\label{subsec:device-mismatch}
In analog hardware, device parameters (such as transistor threshold voltages, transistor sizes, capacitance value, etc) can vary slightly due to manufacturing variations. 
This device mismatch leads to deviations in the behavior of components between identical hardware units~\cite{kinget2005device}.

In the presence of device mismatch, the system dynamics are altered by perturbations $\vecmismatch$ drawn from a probability distribution, typically $\mathcal{N}\left(1, \sigma^2 \cdot \mathbb{I}\right)$, a multivariate normal distribution with unit mean and covariance $\sigma^2 \cdot \mathbb{I}$. For instance, mismatched can cause the ODE system to evolve with perturbed parameters $\vecparam' = \vecmismatch \circ \vecparam$, e.g., Figure~\ref{eq:cnn:dynamics-mismatched}. More generally, mismatches can multiply with any term in the derivative function $f$. Consequently, each realization of the ODE system with a different sample of $\vecmismatch$ follows the dynamics

\begin{equation}
    \frac{d\vecstatevar(t)}{dt} = f_{\vecmismatch}(\vecstatevar(t), \vecparam, t),
\end{equation}
leading to a slightly different trajectory for the $\vecstatevar(t)$ and variation in the loss function $\mathcal{L}_{\vecmismatch}$. It is important to note that $\vecmismatch$ is always a multiplicative term and thus does not affect the differentiability of the system. Hence, backpropagation and the continuous adjoint method remain applicable.

The loss gradient $\frac{d \mathcal{L}}{d \vecparam}$ is calculated using Monte Carlo sampling, where multiple realizations of the mismatch vector $\vecmismatch$ are drawn and the gradient $\frac{d \mathcal{L}_{\vecmismatch}}{d \vecparam}$ is computed for each realization. The final gradient is then obtained by averaging over all the Monte Carlo samples:

{\small
\begin{equation}
    \frac{d \mathcal{L}}{d \vecparam} = \mathbb{E}_{\vecmismatch}\left[ \frac{d \mathcal{L}_{\vecmismatch}}{d \vecparam} \right].
    \label{eq:loss:mismatch}
\end{equation}
}

\subsection{Transient Noise}
When the temperature is above absolute zero, the electrons in the devices move randomly, causing a tiny but unpredictable current which is called thermal noise.
Thermal noise in the time domain is modeled as stochastic processes acting on the state variables over time, and the system dynamics are described by a stochastic differential equation (SDE)~\cite{kidger2022neural,li2020scalable}. Let $g: \mathbb{R}^n\times\mathbb{R}^m\times\mathbb{R} \rightarrow \mathbb{R}^{n\times q}$ be a deterministic function and differentiable with respect to $\vecstatevar$ and $\vecparam$, a SDE is of the form

{\small
\begin{equation}
    d\vecstatevar(t) = f(\vecstatevar(t), \vecparam, t) dt + g(\vecstatevar(t), \vecparam, t) dW(t),
\end{equation}
}

where $W(t)$ represents a $q$-dimensional Wiener process with prescribed correlation.
This equation can be solved using numerical methods such as the Euler-Maruyama scheme, a variant of Euler's method with the update rule:

{\small
\begin{equation}
    \vecstatevar_{t_{i+1}} = \vecstatevar_{t_{i}} + f(\vecstatevar_{t_{i}}, \vecparam, t_{i})\Delta t  + g(\vecstatevar_{t_{i}}, \vecparam, t_{i})\vectransient_{t_i},
\end{equation}}
where $\vectransient_{t_i}$ is an increment of the Wiener process which is essentially a multivariate normal distribution $\mathcal{N}\left(0, \Delta t\cdot\mathbb{I}\right)$.
Each SDE solve requires sampling $\vectransient$ at every time point, leading to solution variations across different runs. Despite this inherent stochasticity, each solve behaves deterministically once a specific sample is drawn, and since the Euler-Maruyama scheme employs only differentiable operators, the process remains fully differentiable with respect to both $\vecparam$ and $\vecstatevar$, enabling backpropagation.

In the presence of transient noise, gradient computation can again be interpreted as a Monte Carlo simulation over noise samples. The gradient $\frac{d \mathcal{L}}{d \vecparam}$ is averaged over samples:

{\small
\begin{equation}
    \frac{d \mathcal{L}}{d \vecparam} = \mathbb{E}_{\vectransient}\left[ \frac{d \mathcal{L}_{\vectransient}}{d \vecparam} \right].
    \label{eq:loss:transient}
\end{equation}}

The continuous adjoint method is also applicable in the SDE setup with some modification~\cite{li2020scalable}.

\proseheading{Transient Noise and Mismatch.}
Since device mismatch is static, solving the system for a perturbation sample is no different from solving the non-mismatched ODE. Thus, incorporating both device mismatch and transient noise in the ODE is straightforward and can be done by adding multiplicative mismatch terms to the derivative ($f$) and the noise amplitude ($g$) functions. The gradient with both mismatch and noise can be computed using Monte Carlo sampling:

{\small
\begin{equation}
    \frac{d \mathcal{L}}{d \vecparam} = \mathbb{E}_{\vecmismatch, \vectransient}\left[ \frac{d \mathcal{L}_{\vecmismatch, \vectransient}}{d \vecparam} \right].
    \label{eq:loss:transientandmismatch}
\end{equation}}

This formulation enables a principled optimization of analog systems in the time domain through gradient descent while ensuring both device mismatch and transient noise are properly addressed.

\subsection{Discrete Parameters}
Analog hardware often implements configurable parameters in a quantized form because the control interface is typically digital, such as switches or digital-to-analog converters (DACs), leading to discrete levels in parameter values. To incorporate such non-differentiable parameters into gradient-based optimization, one effective method is to use the Gumbel-Softmax estimator~\cite{jang2016categorical}. This method approximates the discrete distribution with a differentiable relaxation, enabling standard backpropagation, and is used in neural network architecture search, quantization-aware training, and physics-aware optical neural network~\cite{jang2016categorical, li2022physics, chang2019data}.

Let $\param_d \in \{p_1, p_2, \ldots, p_k\}$ be a discrete parameter. The Gumbel-Softmax estimator introduces a relaxation $\hat{\param_d}\approx\mathbf{y}^\top\mathbf{p}$, where $\mathbf{p}=[p_1, p_2, \ldots, p_k]^\top$ is a vector of discrete levels. $\mathbf{y}$ is a relaxed one-hot vector with each entry calculated as

{\small
\begin{equation}
    y_i = \frac{\exp((\log(\pi_i) + g_i) / \tau)}{\sum_{j=1}^k \exp((\log(\pi_j) + g_j) / \tau)},
\end{equation}}

where $\pi_i$ is a learnable probability for the discrete level $p_i$, i.e., the larger the $\pi_i$, the more likely that $\hat{\param_d}\approx p_i$. $g_i$ is a small perturbation drawn from $\text{Gumbel}(0, 1)$ distribution~\cite{jang2016categorical}. The temperature parameter $\tau$ controls the smoothness of the approximation: as $\tau \to \infty$, the distribution approaches a uniform distribution, while as $\tau \to 0$, $\mathbf{y}$ becomes identical to a one-hot vector. Because $\frac{\partial \mathbf{y}}{\partial \boldsymbol{\pi}}$ is well-defined, replacing non-differentiable discrete parameters with the Gumbel-Softmax estimator allows for the use of backpropagation.
In practice, $\tau$ is annealed from high value to small but nonzero value with a linear schedule or an exponential decay schedule.

\subsection{Bounded Parameters}
In analog hardware, the parameter magnitudes vary widely due to differences in their physical characteristics. For example, capacitances may be at the pico scale and currents at the micro-scale. Therefore, normalization is critical to ensure stable and efficient gradient descent-based optimization. In practice, these parameters fall within physically reasonable ranges, allowing for normalization with the physical range followed by a clipping within certain limits, such as $[-1, 1]$.

\section{Implementation of the \tool{} Framework}
\label{sec:framework}
The \tool{} framework leverages the JAX ecosystem for implementing the differentiable analog system models with nonidealities~\cite{jax2018github}. Specifically, we utilize Diffrax, a JAX-based library specialized in solving differential equations that supports both backpropagation through the solver and the continuous adjoint method for ODEs and SDEs~\cite{kidger2022neural}. 

\tool{}'s analog modeling APIs are built on top of Ark, a programming language designed for defining analog compute paradigms~\cite{wang2024design}. Ark provides an expressive syntax for specifying analog compute models, which can be compiled into ODEs. By combining JAX’s high-performance automatic differentiation with Ark's flexibility, \tool{} enables seamless optimization of complex analog systems.

\subsection{Frontend APIs}

\tool{} offers several core APIs to manage various aspects of analog systems, including noise, device mismatch, and discrete variables. These APIs integrate automatically with Diffrax’s ODE/SDE solvers and JAX’s autograd features:
\begin{itemize}
    \item \codein{AnalogTrainable} and \codein{DiscreteTrainable}: The two classes represent trainable continuous and discrete parameters that will resolve to targets for gradient descent update.
    
    \item \codein{mismatch}: This decorator introduces device mismatch into a trainable/non-trainable parameter. It adds a multiplicative symbolic mismatch term, which is concretized in execution time by drawing samples from a normal distribution with the provided relative standard deviation.

    \item \codein{noiseexp}: This API allows for the specification of a noise amplitude function similar to defining a derivative function in Ark and \tool{} automatically adds to the system's ODEs to form SDEs, as Ark does not provide SDE support.
\end{itemize}

\subsection{Compilation to Differentiable Functions}

Based on the provided specifications, \tool{} compiles the model into a differentiable function within the JAX framework. The forward pass solves the system of ODEs (or SDEs), while the backward pass computes gradients of the loss function with respect to the trainable parameters, using either backpropagation or the continuous adjoint method.

Once compiled as a differentiable function, the analog system can be optimized similarly to optimize a neural network. Users provide input-output data, random seeds for Monte Carlo sampling, and a loss function based on the system’s output versus the target output. Then, users can apply standard deep learning techniques for optimization, such as using the Adam optimize for gradient descent, with GPU acceleration enabling large-scale parallelizationr~\cite{kingma2017adam}.


\subsection{Discussion}

\proseheading{Modeling of Analog Hardware.} \tool{} works with the differential equation representation of the underlying analog hardware, following the same design philosophy as Ark~\cite{wang2024design}. \tool{}'s ODE models operate at the same level of abstraction as behavioral, Simulink, and equivalent circuit models used in the early stages of circuit design. The parameterized ODEs produced by \tool{} serve as an executable requirements specification, constraining and informing the transistor-level unit cell designs, which would be developed with a fabrication-ready PDK. With this modeling approach, the ODE models are expected to be iteratively refined to capture more non-idealities at a finer granularity as the transistor-level designs are developed. The parameter constraints limit the ODE model to only allow reasonable parameterizations.

\section{Case Studies}
We use \tool{} in two design case studies to maximize an end-to-end application-specific quality metric. The case study statistics and optimizer runtimes are summarized in Table~\ref{tab:benchmark-characterstic}. To our knowledge, this is the first work demonstrating the optimization of these systems in the presence of non-idealities. All experiments were conducted on an Intel(R) Xeon(R) Silver 4216 CPU paired with a Quadro RTX 6000 GPU.

\begin{table*}[t]
\footnotesize
\begin{tabular}{|c|c|c|cc|c|c|c|}
\hline
\multirow{2}{*}{\begin{tabular}[c]{@{}c@{}}Compute\\ Paradigm\end{tabular}} &
  \multirow{2}{*}{Application} &
  \multirow{2}{*}{\begin{tabular}[c]{@{}c@{}}Modeled Hardware\\ Characteristics\end{tabular}} &
  \multicolumn{2}{c|}{Trainable Paramters} &
  \multirow{2}{*}{\begin{tabular}[c]{@{}c@{}}\# of state\\ variables\end{tabular}} &
  \multirow{2}{*}{\begin{tabular}[c]{@{}c@{}}Cost \\ function\end{tabular}} &
  \multirow{2}{*}{\begin{tabular}[c]{@{}c@{}}Optimization\\  runtime\end{tabular}} \\ \cline{4-5}
 &
   &
   &
  \multicolumn{1}{c|}{physical meaning} &
  \# &
   &
   &
   \\ \hline
\cnn &
  \begin{tabular}[c]{@{}c@{}}Edge \\ detection\end{tabular} &
  Device mismatch &
  \multicolumn{1}{c|}{\begin{tabular}[c]{@{}c@{}}Feedforward, feedback\\ and bias magnitude\end{tabular}} &
  19 &
  256 &
  MSE &
  33 mins \\ \hline
\obc &
  \begin{tabular}[c]{@{}c@{}}Pattern\\ recognition\end{tabular} &
  \begin{tabular}[c]{@{}c@{}}Discrete parameters\\ Transient noise\end{tabular} &
  \multicolumn{1}{c|}{\begin{tabular}[c]{@{}c@{}}Coupling and \\ locking weight\end{tabular}} &
  104 (+1) &
  60 &
  MSE &
  4 mins \\ \hline
\tln &
  PUF &
  Device mismatch &
  \multicolumn{1}{c|}{\begin{tabular}[c]{@{}c@{}}Capacitance, inductance,\\ transconductance values\end{tabular}} &
  25 &
  514 &
  $\itoo$ &
  1.4 hrs \\ \hline
\end{tabular}
\caption{Benchmark Characteristics and Optimizer Runtime.}
\label{tab:benchmark-characterstic}
\end{table*}

\subsection{Oscillator-Based Pattern Recognizer}\label{subsec:example:obc}

In recent years, \obc{} has been used to solve combinatorial optimization problems energy efficiently and perform pattern recognition~\cite{chou2019Con, ochs2021Con, mallick2021con-global,kumar2017autoassociative, todri2021frequency, maffezzoni2015oscillator}. In oscillator-based computing, information is encoded in the phase of an oscillating signal instead of traditional voltage or current levels. To perform computation, oscillators are coupled together; coupled oscillators will tend to synchronize so that they are either in-phase or out-of-phase.  The ideal phase dynamics of the oscillators in \obc{} are captured with the \textit{Kuramoto model}, a nonlinear differential equation~\cite{ochs2021Con, wang2017oscillator}:

{\small
\begin{equation}\label{eq:obc}
    \ddt{x_i} = -\sum\limits_{j=1}^{n}\cplweight\sin{(\pi(x_i-x_j))} - \lockweight\sin{(2\pi x_i)}, i=1,...,n. 
\end{equation}
}

The \(x_i\) state variables model the phase of oscillator $i$ over time normalized by $\pi$, while the \(\cplweight\cdot\sin{(\pi(x_i-x_j))}\) term characterizes the interaction between oscillators \(i\) and \(j\) with coupling weight \(\cplweight\). On measurement, we center the phase to lie between $[0,1]$. The more positive the coupling weight, the stronger the tendency for the two oscillators to synchronize in phase. Conversely, the more negative the coupling weight, the stronger the tendency for the oscillators to settle into distinct phases. The injection locking weight $\lockweight$ controls the convergence speed~\cite{ochs2021Con, wang2017oscillator,maffezzoni2015oscillator}.

\begin{figure}[t]
    \centering

    \begin{minipage}{0.44\linewidth}

    \subfloat[\txsfig{(Random,Couple)}, Initial, $\mathcal{L}$=0.244]{
        \centering
        \includegraphics[width=\linewidth]{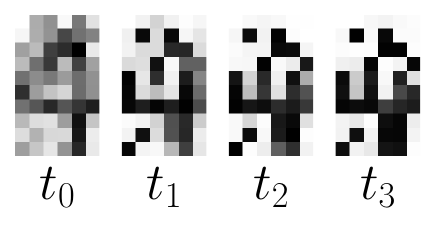}
        \label{fig:obc:evol:1-bit-rand}
    }

        \subfloat[\txsfig{(Hebbian,-)}, Initial, $\mathcal{L}$=0.156]{
        \centering
        \includegraphics[width=\linewidth]{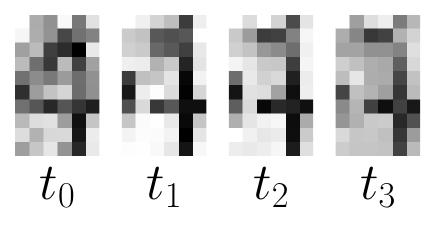}
        \label{fig:obc:evol:1-bit-hebb}
    }

    \subfloat[\txsfig{(Random,Couple)}, Optimized, $\mathcal{L}$=0.073]{
        \centering
        \includegraphics[width=\linewidth]{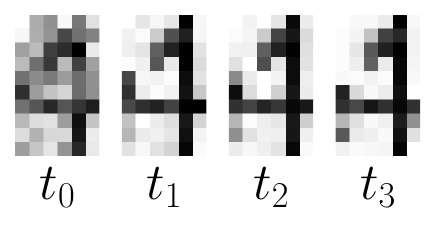}
        \label{fig:obc:evol:1-bit-opt}
    }
    
    \subfloat[\txsfig{(Hebbian,Couple\&Lock)}, Optimized, $\mathcal{L}$=0.017]{
        \centering
        \includegraphics[width=\linewidth]{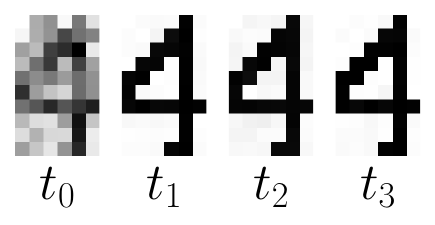}
        \label{fig:obc:evol:1-bit-opt-locking}
    }

    \end{minipage}\hfill%
    \begin{minipage}{0.55\linewidth}

    \centering
    \subfloat[Stored Patterns]{
        \includegraphics[width=0.6\linewidth]{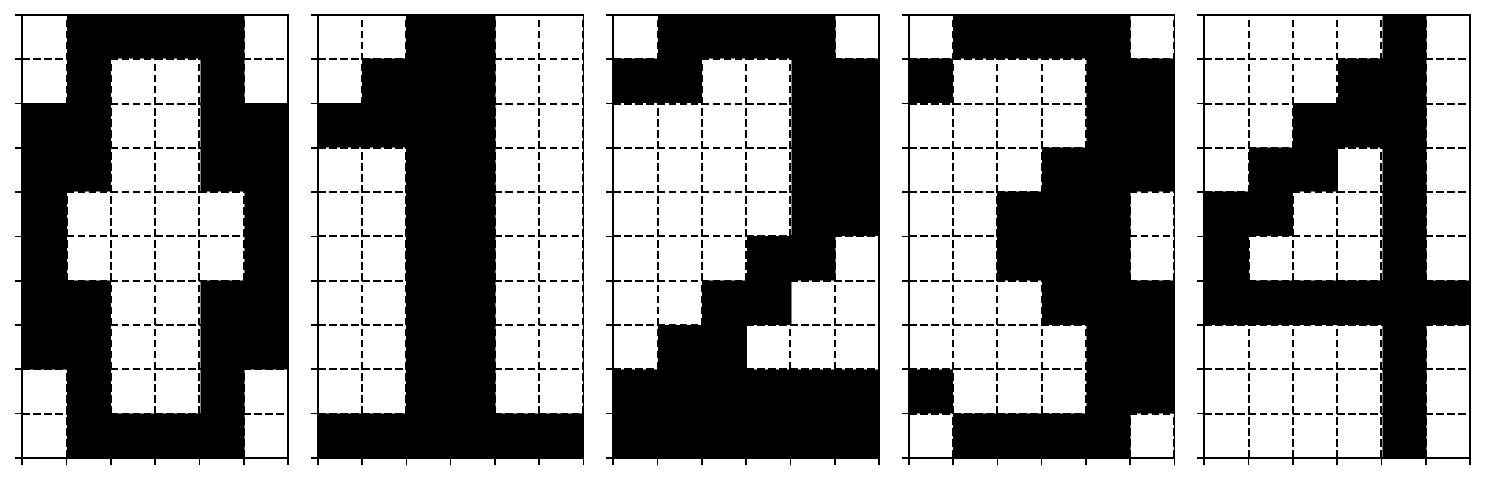}
        \label{fig:obc:patterns}
    }
    
    \subfloat[\txsfig{(Random,Couple)}]{
        \centering
        \includegraphics[width=0.75\linewidth]{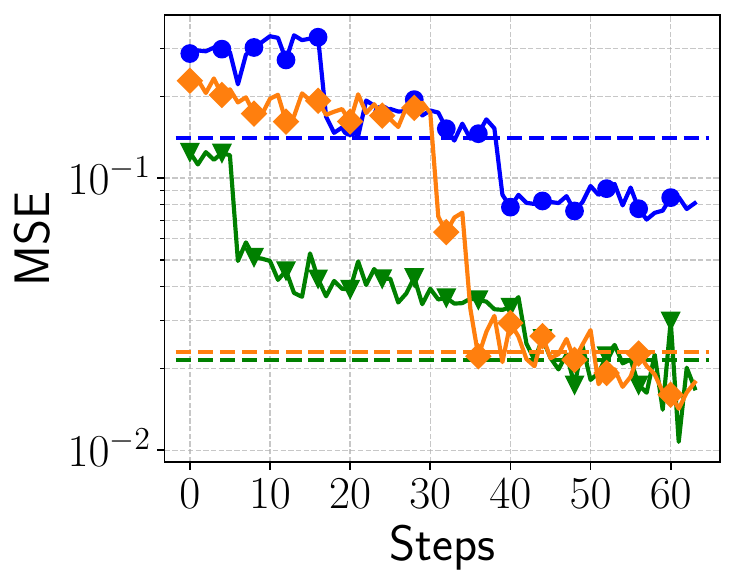}
        \label{fig:obc-loss-vs-iter-rand-init}
    }

    \subfloat[\txsfig{(Random,Couple\&Lock)}]{
        \centering
        \includegraphics[width=0.75\linewidth]{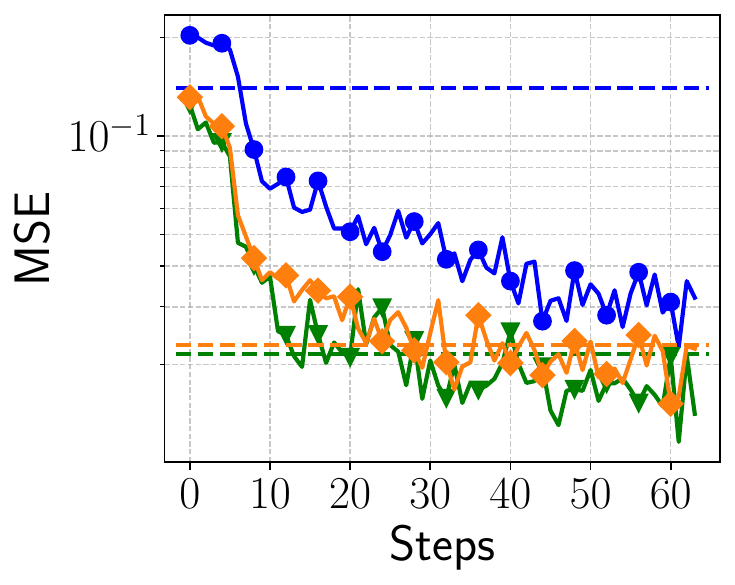}
        \label{fig:obc-loss-vs-iter-rand-init-train-locking}
    }
    
    \subfloat[\txsfig{(Hebbian,Couple\&Lock)}]{
        \centering
        \includegraphics[width=0.75\linewidth]{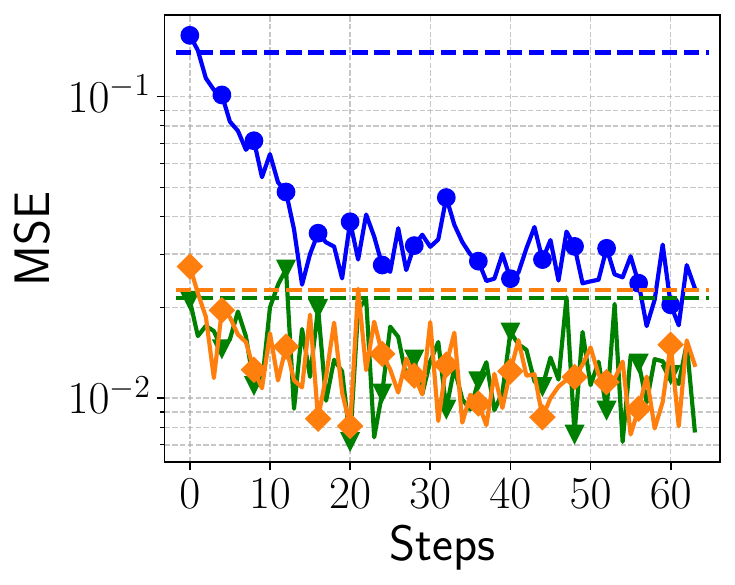}
        \label{fig:obc-loss-vs-iter-hebb-init-train-locking}
    }\hfill

    \end{minipage}

        \small
    \subfloat[MSE Loss Comparison after Optimization]{
    \footnotesize
    \begin{tabular}{|cc|c|c|c|c|}
        \hline
        \multicolumn{2}{|c|}{Initial Coupling} & Hebbian & Random & Random & Hebbian \\ \hline
        \multicolumn{2}{|c|}{Trainables} & - & \txsfig{Couple} & \begin{tabular}[c]{@{}c@{}}\txsfig{Couple} \\ \txsfig{\& Lock}\end{tabular} & \begin{tabular}[c]{@{}c@{}}\txsfig{Couple} \\ \txsfig{\& Lock}\end{tabular} \\ \hline
        \multicolumn{1}{|c|}{\multirow{3}{*}{\begin{tabular}[c]{@{}c@{}}\txsfig{Couple} \\ bit width\end{tabular}}} & 1 & 0.140 & 0.077 & 0.029 & 0.023 \\ \cline{2-6} 
        \multicolumn{1}{|c|}{} & 2 & 0.022 & 0.019 & 0.017 & 0.011 \\ \cline{2-6} 
        \multicolumn{1}{|c|}{} & 3 & 0.023 & 0.017 & 0.019 & 0.010 \\ \hline
    \end{tabular}
    \label{fig:obc:testing}
    }\\
    
    \caption{\obc{} Pattern Recognizer. \textit{Left:} dynamics of \obc{} with 1-bit weights on noisy "4" input. \textit{Right:} Loss \swatch{blue}: 1-bit; \swatch{matplotlibgreen}: 2-bits; \swatch{orange}: 3-bits, y-axis: log(MSE), x-axis: iteration, dotted: Hebbian rule loss. \textit{Bottom.} Loss for different optimizer setups.}
    \label{fig:obc:evol}
\end{figure}

\proseheading{Pattern Recognition.}
Networks of oscillators can “memorize” patterns and recover them from noisy inputs~\cite{kumar2017autoassociative, todri2021frequency, maffezzoni2015oscillator}. We parameterize the network to memorize five digits, each represented as a 10$\times$6 pixel grid with black (0) and white (1) pixels, where each pixel $p_i$ is mapped to an oscillator $x_i$ (Figure~\ref{fig:obc:patterns}). The patterns are programmed into the \obc{} by setting coupling weights $\allcplweight$, and the locking weight $\lockweight$ is determined through parameter sweeping in simulation~\cite{todri2021frequency}.

The goal of the recognizer is to have the oscillator phases converge to the original pattern from a noisy greyscale image, where each pixel is in $\left[0,1\right]$ (e.g., Figure~\ref{fig:obc:evol:1-bit-opt-locking}).  Each oscillator's phase $\phi$ is directly set to the corresponding noisy image's pixel value $v_i$. The dynamical system then evolves over time and settles to the memorized pattern that is the closest match. To recover the rectified image from the oscillators' phase, we simply measure the phase of each oscillator.




\subsubsection{Optimization with Shem}

In this case study, we use \tool{} to learn the coupling weights from the dataset while simultaneously considering the nonidealities and resource limitations in the hardware implementation. We compare the efficacy of the learned hardware-aware coupling weights against a Hebbian rule-based approach, which deterministically computes coupling weights from the patterns to memorize and tunes the locking weight without considering nonidealities~\cite{kumar2017autoassociative, todri2021frequency, maffezzoni2015oscillator}. We describe the target hardware and then present our optimization results.

\subsubsection{\obc{} Hardware}\label{subsec:example:obc:hw}

Hardware implementations of \obc{} are susceptible to transient phase noise, which causes deviations in oscillator dynamics captured in the Kuromoto model in Equation~\ref{eq:obc}, potentially degrading pattern recognition accuracy~\cite{lee2000oscillator}. This is typically modeled as white noise in the Kuramoto model~\cite{wang2017oscillator}. In this study, we introduce phase noise with a standard deviation of $\alpha=0.025$.

{\small
\begin{equation}\label{eq:obc-noise}
\ddt{x_i} = -\sum\limits_{j=1}^{n}\cplweight\sin{(\pi(x_i-x_j))} - \lockweight\sin{(2\pi x_i)} + \textcolor{blue}{\noiseamp\xi_i(t)},
\end{equation}}

where $\xi_i(t)$ is a Gaussian white noise processe and $\noiseamp$ is the noise amplitude. While small relative to coupling and locking signals, noise introduces random phase perturbations that impair synchronization. Noise is typically not considered when selecting the coupling weights.

\proseheading{Limited Connectivity.} We arrange the oscillators in a grid and limit our \obc{} hardware to only allow neighboring oscillators to be coupled together. We limit coupling to neighbors because supporting all-to-all coupling between oscillators significantly increases area usage. For example, within the same device technology and chip area, an all-to-all connection accommodates 30 oscillators, with most of the area consumed by routing~\cite{mallick2021con-global}, while a neighboring connection supports up to 560 oscillators~\cite{ahmed2021con-neighbor}. Thus, for large-scale problems, a neighboring connection is preferable. Hardware connectivity limitations are typically not considered when designing the \obc{} application, and all-to-all connectivity is expected.

\proseheading{Discrete Coupling Weight.}
In practice, the coupling weights are digitally programmed with a digital-to-analog converter (DAC), where the DAC's bit width controls its precision~\cite{moy20221ringosc}. Because higher-precision DACs consume more area and energy, selecting the DAC with the smallest bit width is preferable. We note that typically, the effects of coupling weight discretization on the computation are not considered.





\subsubsection{Evaluation}
We use \tool{} to optimize the \obc{} paradigm for the pattern recognizer under transient noise, with trainable discrete coupling weights in a neighboring connection topology. We use a dataset containing 512 pairs of digit images, where each pair contains an ideal and a noisy image with $\text{uniform}(-0.5, 0.5)$ noise added. The loss is the MSE between the \obc{} image readout and ideal image at measurement time $t=1$.  Both the noisy image and transient noise are resampled at the start of each batch to avoid overfitting. 

We execute the Adam optimizer with a learning rate of 0.1 for 64 steps and anneal the temperature of the Gumbel softmax exponentially from 10 to 1. We report the MSE of the lowest-training-loss parameter set on a test set of 8192 image pairs. The Hebbian baseline quantizes the coupling weights obtained from the Hebbian rules to the nearest discrete value and discards all non-neighboring oscillator weights.


\proseheading{Analysis.}
We initially learned coupling weights for 1, 2, and 3-bit DACs with the locking weight initialized to one. We use \txs{(Initialization, Trainables)} to denote each instantiation; the \txs{(Random, Couple)} run starts from a random initialization and optimizes the coupling weights, for example. Figure~\ref{fig:obc:evol:1-bit-rand} and~\ref{fig:obc:evol:1-bit-hebb}  shows that both the initial random parameterization and Hebbian baseline struggle to recover the "4" digit, with the Hebbian baseline doing marginally better.

Figures~\ref{fig:obc-loss-vs-iter-rand-init}, \ref{fig:obc-loss-vs-iter-rand-init-train-locking}, and \ref{fig:obc-loss-vs-iter-hebb-init-train-locking} track the MSE loss over 60 iterations for each instantiation. The loss steadily decreases for both \txs{Random} and \txs{Hebbian} as the optimizer finds better solutions.
Figure~\ref{fig:obc:testing} illustrates the MSE loss for different optimization setups, where learned coupling and locking parameters achieve MSE reductions of 0.03-0.117 compared to Hebbian rule initialization. 
After optimizing the coupling weights with a 1-bit DAC, the results shown in Figure~\ref{fig:obc:evol:1-bit-opt} indicate an improvement, as the \obc{} recovers a greater portion of the "4" digit, though the solution remains suboptimal.

Introducing trainable locking yields better results, especially in the 1-bit DAC configuration, with the loss reduced by 0.048 compared to the coupling-only setup. However, for 2-bit and 3-bit DACs, the optimizer converges to similar losses. The \txs{(Hebbian, Couple\&Lock)} setup achieves the best performance across all configurations, indicating a good initial guess leads to better convergence, which is common in gradient-based optimization. Notably, the 1-bit optimized coupling and locking run achieves a comparable loss to the 3-bit Hebbian initialization. Figure~\ref{fig:obc:evol:1-bit-opt-locking} shows qualitatively that the 1-bit \obc{} can recover "4" with a 1-pixel error in a noisy setting. We therefore identify the following design tradeoff: For a resource-bounded system, 1-bit DACs are preferred with acceptable loss; Otherwise, 2-bit DACs offer lower loss. There is no advantage in using 3-bit DACs, as the loss doesn't improve.

\subsection{Transmission Line-Based Security Primitive}

Transmission line-based computing (TLN) is an analog compute model that mimics the propagation of electromagnetic waves through transmission lines to perform computation~\cite{wang2024design}. The transmission lines are organized into a network, where the reflection, transmission, and superposition of waves are used to perform computation. The system is modeled by the Telegrapher’s equations, discretized for the line segments:

\begin{equation}
    \begin{cases}
        \frac{dV_i}{dt} &= \frac{1}{C_i}(I_i - I_{i+1}), \\
        \frac{dI_i}{dt} &= \frac{1}{L_i}(V_{i-1} - V_i),
    \end{cases}
\end{equation}

Here, \( V_i \) and \( I_i \) are the voltage and current at segment \( i \), and \ \( L_i \), and \( C_i \) are the inductance and capacitance parameters, respectively, and control propagation speed. With this paradigm, we are primarily interested in measuring the traveling wave on a particular link at the readout time \textit{before} the system has settled. In contrast, most analog computing models only use the steady-state value.

\proseheading{TLN Hardware.} The TLN model can be emulated with analog circuits typically used to implement filters. The hardware implementation introduces the scaling parameters \( gc_{t,i} \), \( gc_{s,i} \), \( gl_{t,i} \), and \( gl_{s,i} \) which map to lumped transistor parameters:

\begin{equation}
    \begin{cases}
        \frac{dV_i}{dt} &= \frac{1}{C_i}(gc_{t,i} \cdot I_i - gc_{s,i+1} \cdot I_{i+1}), \\
        \frac{dI_i}{dt} &= \frac{1}{L_i}(gl_{t,i-1} \cdot V_{i-1} - gl_{s,i} \cdot V_i). 
    \end{cases}
\end{equation}

Mismatch in these parameters causes deviations from the expected behavior, affecting the system's transient dynamics.

\begin{figure}[t]
\subfloat[Switchable-Star PUF (SS-PUF) Topology]{
    \centering
    \includegraphics[width=0.6\linewidth, valign=b]{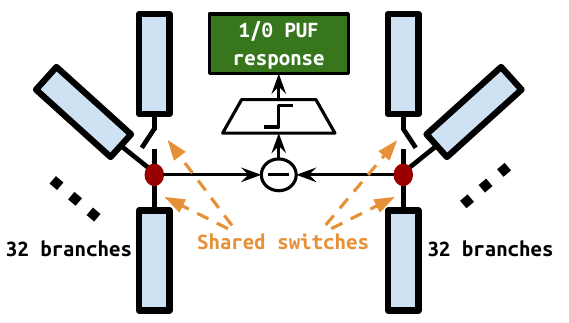}
    \label{fig:tln:puf:topology}
}\hfill%
\adjustbox{valign=b}{
        \subfloat[I2O Loss Values]{
        \footnotesize
        \begin{tabular}{|c|c|}
        \hline
        Parameters & I20 Loss \\
        \hline
        Initial & 0.130\\
        \hline
        Optimized & 0.084\\
        \hline
        Optimized, & \multirow{2}{*}{0.119}\\
        Fixed $L_0C_0$ &\\
        \hline
        \end{tabular}
    \label{fig:tln:testing}
    }
}

\subfloat[SS-PUF Simulation]{
    \includegraphics[width=0.45\linewidth]{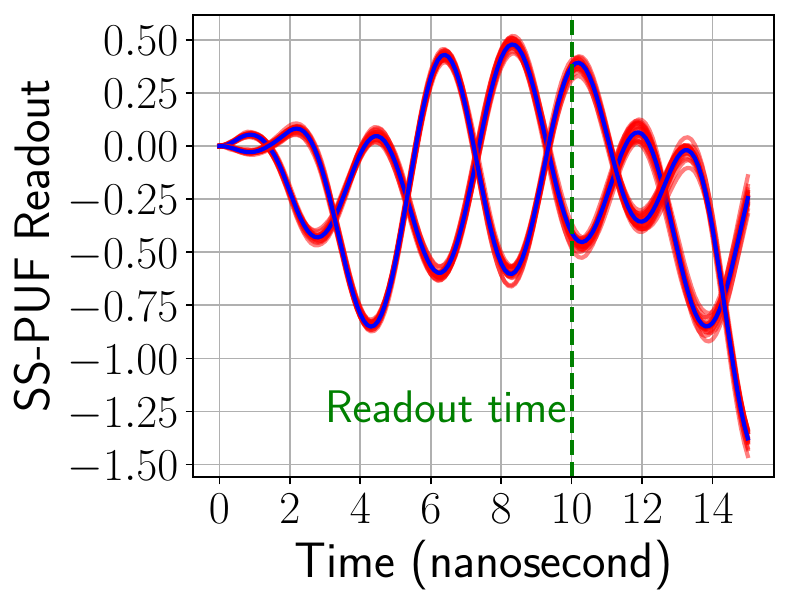}
    \label{fig:trajectory-mismatch-transient}
}\hfill%
\subfloat[$\itoo$ Loss over Optimizer Steps]{
        \centering
        \includegraphics[width=0.45\linewidth]{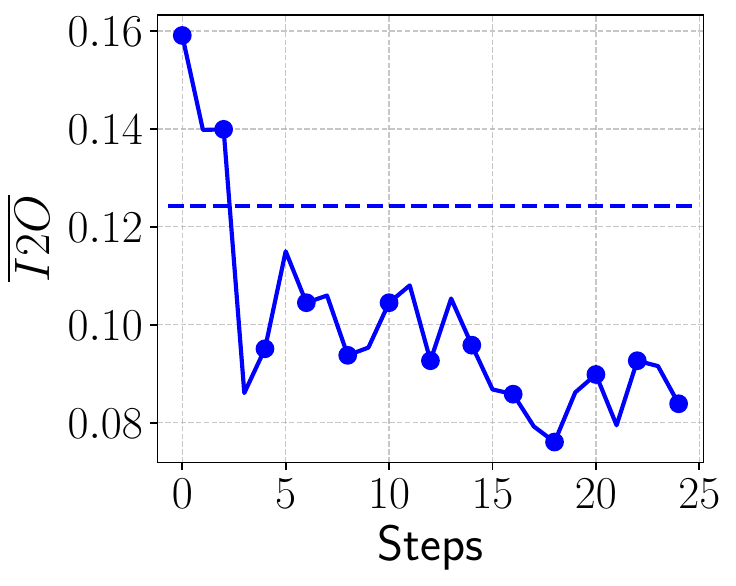}
        \label{fig:tln-loss-vs-iter}
}

\caption{Transmission Line PUF. \textit{Top-left.} Switchable-Star PUF. \textit{Bottom-left.} Simulation of two mismatched designs: \swatch{blue}: mismatch only, \swatch{red}: mismatch and noise. \textit{Right.} optimizer loss }
\end{figure}

\subsubsection{\tln{}-based PUF}

A Physical Unclonable Function (PUF) is a hardware security primitive that leverages intrinsic manufacturing variations to generate unique and unpredictable responses (output bits) for given challenges (input bits). For this application, we exploit rather than eliminate the effect of process variation and optimize the system to amplify the effect of mismatch on the dynamics of the TLN~\cite{wang2024design, csaba2009cnn-puf}. In this case study, we start with a 32-bit "Switchable Star" TLN PUF design (SS-PUF), pictured in Figure~\ref{fig:tln:puf:topology}. The SS-PUF has two nominally identical circuits (without mismatch), each with 32 switchable branches.  Each branch is a transmission line controlled by a programmable switch that connects or disconnects the branch, where the switch is set by a challenge bit. During execution, a pulse is dispatched at the stars' centers, and the response is measured by thresholding the signal difference of the center nodes at $t=10$ nanoseconds. A positive signal is mapped to a "1" response bit. We use a logistic function to make the thresholding operation on readout differentiable.

\proseheading{Evaluation Metric.} We optimize the $\itoo$ (mean instance-to-optimum) metric, a PUF security metric that measures the bias of the response bit over the space of similar challenges~\cite{stephani2024i2o}. The intuition is that the response to a given challenge should not reveal information about the response to a slightly modified challenge; otherwise, a savvy attacker can exploit these correlations. Formally, $\itoo$ for $r$ instances of a PUF design with $n$-bit challenges is defined as:

\begin{equation}
    \itoo= \frac{1}{r} \sum_{i=1}^{r} \left( \frac{1}{n} \sum_{j=1}^{n} \left| S_j(P_i) - 0.5 \right| \right),
    \label{eq:tln:i2o}
\end{equation}

where $S_j(P_i)$ represents the probability that flipping the $j$-th bit of a challenge for the $i$-th instance results in a flipped response, and $n$ is the number of challenge bits. We observe this PUF metric is differentiable.





\subsubsection{Optimization with \tool{}}

We use \tool{} to optimize the $gc$, $gl$, $L$, and $C$ parameters in the SS-PUF to minimize the $\itoo$ loss. To the best of our knowledge, this is the first work to optimize PUF design parameters to minimize/maximize a security metric.

\noindent\textit{Experimental Setup.} For the optimization process, we use eight mismatched parameter sets and 32 sets of similar challenges per batch, where each set contains 32 challenges that are one bit flip apart from a ``reference'' challenge. Both the mismatches and challenges are resampled in every batch to avoid overfitting. The system is optimized using the Adam optimizer with a learning rate of $0.005$ for a total of 24 steps.
All parameters are \codein{AnalogTrainable} with initialization $C_i=L_i=10^{-9}$ and $gc_{t,i}=gc_{s,i}=gl_{t,i}=gl_{s,i}=1$, which produce the baseline $\itoo=0.130$. We constrain each link in the SS-PUF to have the same parameter values; we learned from designing the SS-PUF that the response is more uniformly sensitive to mismatch if symmetry is preserved. For testing, we evaluate 192 mismatched PUFs with the same challenge setup as during training.

\proseheading{Analysis.} Figure~\ref{fig:tln-loss-vs-iter} shows the $\itoo$ loss over optimization iterations. Fluctuations in the loss curve are expected due to the limited subspace of the PUF that is sampled during optimization ($2^{10}$ out of the total $2^{32}$ challenge space). Despite these fluctuations, a clear overall trend of decreasing loss is visible. This improvement remains consistent in testing, improving $\itoo$ from 0.130 to 0.089. Therefore, the optimized design improves statistical security properties of the design.


The optimized ODE produces TLN trajectories very sensitive to mismatch, but not chaotic. Figure~\ref{fig:trajectory-mismatch-transient} shows the noiseless (blue lines) and noisy (red lines) response of two mismatched SS-PUFs instances, where we inject white noise with a standard deviation of $1 \times 10^{-7}$ estimated with a MOS transistor thermal noise model~\cite{Razavi2005analog-design}. We observe the noisy trajectories closely follow the noiseless trajectory for the same instance, and the two mismatched instances produce very different trajectories. In a chaotic system, a very small perturbation from noise would produce a drastically different result in a chaotic system, rendering the PUF unstable. 



With \tool{}, we can easily study the importance of certain parameters. We perform multiple optimization runs and observe one of the runs has difficulty reducing the loss. We compare the parameters and find that the failing run chooses very different center node capacitance $C_0$ and inductance $L_0$ values from the succeeding runs. We hypothesize the center node parameters have an outsized effect on the $\itoo$ score and optimize the SS-PUF with non-trainable $C_0$, $L_0$. With these two parameters fixed, \tool{} achieves a loss of $0.119$ compared to $0.084$, indicating the $C_0$, $L_0$ parameters have an outsized effect on the security properties of the PUF.

\section{Related Work}

\noindent\textbf{Analog EDA Tools.} 
The “adjoint sensitivity analysis” technique in analog design automation originated in the late 1960s, and has since been expanded by numerous studies, including~\cite{rohrer1967fully,director1969generalized, conn1998jiffytune, conn1999gradient, visweswariah2000noise, joshi2017analog}. These works focus on device-level optimization, specializing in circuit-specific metrics such as delay or gain. Our approach works with higher-level ODE models, optimizing end-to-end system-level cost functions. Additionally, we support nonidealities and digital control logic, which is crucial for mixed-signal design.

\noindent\textbf{Adjoint Method for Design Optimization.}
The adjoint method is extensively used in the inverse design, as discussed in~\cite{givoli2021tutorial, lettermann2024tutorial}. For example, in the inverse design of photonic devices, they perform a forward simulation of the Maxwell's equation with the finite-difference frequency-domain (FDFD) or finite-difference time-domain (FDTD) method and solve the adjoint with a backward simulation~\cite{elesin2014time, hughes2018adjoint, su2020nanophotonic, molesky2018inverse}. Their framework targets Maxwell's equations, a specific partial differential equation model, to model the electromagnetic dynamics of photonic devices. We model general ordinary differential equations with first-class support of nonidealities common in unconventional analog systems. 

\noindent\textbf{Physics-Aware Training.}
Practitioners have developed physics-aware training algorithms for ML pipelines used for in-sensor compression and machine vision, diffractive optical neural networks for image classification, and deep physical neural network~\cite{wright2022deep, ma2023leca,li2022physics, li2023lightridge}. They implement some or all parts of the neural network with their hardware platform and utilize differentiable hardware models to include physical behaviors or nonidealities during training. The modeling process is sophisticated and manually done in a per-hardware manner. Our work complements this area of research by introducing a framework that enables a systematic description of analog computing paradigms and their compilation into differentiable models, simplifying the modeling process in physics-aware training. Moreover, our framework extends beyond neural network training and supports the general optimization of unconventional analog systems.

\noindent\textbf{Physics-Informed Neural Network}
A physics-informed neural network (PINN) integrates physical laws into its loss function to guide the learning process and has applications in solving forward and inverse problems in fluid dynamics, geoscience, material design, etc.~\cite{cai2021physics,cuomo2022scientific}. In contrast, our approach directly simulates systems, ensuring that the underlying dynamics are faithfully captured by construction rather than approximated by a trained neural network. These methods are complementary -- when parts of an analog system are too complex to model directly, they can be represented using a PINN, and since both approaches are differentiable, the entire system can be optimized end-to-end.

\noindent\textbf{Neural Differential Equations.}
Neural differential equations (NDEs) use neural networks to model state variable derivatives and have applications in generative models, dynamical systems, and time series prediction~\cite{chen2018neural, kidger2022neural, li2020scalable}. In contrast, our approach directly computes the physical dynamics in analog computations. Interestingly, analog computing, which naturally solves differential equations, is well-suited for NDE hardware, as suggested by~\cite{watfa2023adjoint}. Our framework provides a flexible tool for exploring and designing analog circuits to support NDE implementations.

\section{Conclusion}

We presented \tool{}, an optimization framework for analog systems with support for key analog hardware characteristics, including device mismatch, noise, and discreteness. \tool{} leverages differentiation methods recently popularized to train neural ODEs to enable optimization of analog systems and is built on the \jax{} ML framework, inheriting the programming conveniences and performance optimizations from the library. We demonstrated that \tool{} optimizes unconventional analog systems in the presence of hardware nonidealities that have not been optimized with automated methods previously.  We view \tool{} as the initial step to connecting the two domains, enabling the application of optimization and gradient-based methods typically reserved for training neural ODEs to optimize analog hardware.

\bibliographystyle{plain}
\bibliography{references}

\end{document}